# The Effect of Extended Cornell Potential on Heavy and Heavy-Light Meson Masses Using Series Method

M. Abu-shady[1] and H. M. Fath-Allah[2]

Department of Mathematics and Computer Sciences, Faculty of Science, Menoufia University, Egypt[1]

Physics and Mathematics Engineering, High Institute of engineering, 15 May, Helwan, Egypt[2]

## Abstract

The effect of an extended Cornell potential on mass spectra of heavy and heavy-light mesons is studied. The Cornell potential is extended to include quadratic potential and inverse quadratic potential. The N-radial Schrödinger equation is solved by using series method. The results for charmonium and bottomonium, and light-heavy meson masses are obtained. A comparison with other recent works is discussed. The present results are improved in comparison with other recent works and are in good agreement with experimental data.

## 1-Introduction

The quantum chromodynamics (QCD) is the fundamental theory of strong interactions that is one of the four fundamental forces of nature. It describes the interactions of quarks, via their color quantum numbers. The study of quarkonium properties plays a great role in analysis the strong interaction of quarks [1 − 4]. The Schrodinger equation (SE) is a key in different branches of science and its solution plays a great role in studying the properties of quarkonium such as the mass spectra of quarkonium by using potential models [ 5 − 6]. Most of researchers have calculated the solution of SE by using different methods such as the numerical methods [7 − 9], the Laplace transform method [10 − 11], The

Nikiforov-Uvarov (UV) method [12 − 14]. The study of quarkonium in the higher dimensional takes much attention in recent works as [15 − 19], which give a lot of the information about nature of interquark forces. The different types of potential model are used such as a mix between the Cornell potential, harmonic potential oscillator potential or / and inverse harmonic potential [20 − 23]. The Cornell potential includes two terms, the Coulomb and linear terms. In this work, the following potential is employed as in Ref. [5]

$$V(r) = ar^2 + br - \frac{c}{r} + \frac{d}{r^2}, \tag{1}$$

where a, b, c and d are positive potential parameters, which will be fixed by considering experimental data later on.

The aims of this work: The first aim, we treat the difficulty that found in Ref. [6] which used the present method. The second aim, we extend the study to other heavy-light mesons that are not calculated in Ref. [6] and other works. Third aim, the effect of dimensionality number on quarkonium mass is investigated. Thus, the N-dimensional of radial SE is analytically solved to obtain the energy eigenvalues and corresponding wave functions by using power series technique.

The paper is organized as follows: In section 2, the power series technique and the energy eigenvalues are calculated in the N-dimensional form. In section 3, the masses spectra of heavy and heavy-light quarkonium are calculated. In section 4, the results are discussed. In section 5, the summary and conclusion are presented.

## 2. The solution of the Schrödinger equation for a given potential using the power series technique

In N-dimensional Hilbert space, the SE for two particles interacting via spherically symmetric potential (1) can be written as in [6]

$$\left[\frac{d^2}{dr^2} + \frac{N-1}{r}\frac{d}{dr} - \frac{l(l+N-2)}{r^2} + 2\mu\left(E - ar^2 - br + \frac{c}{r} - \frac{d}{r^2}\right)\right]R(r) = 0, \qquad (2)$$

where $l$ is the angular momentum quantum number, N denotes dimensionality and $\mu = \frac{m_1 m_2}{m_1+m_2}$ is the reduced mass of the particles of masses $m_1$ and $m_2$, E are the energy eigenvalues corresponding to the radial eigenfunctions. R(r) is wave function. Now, we find an approximate solution to equation (2) by making the following choice the wave function as in Ref. [6]:

$$R(r) = \exp(-\alpha r^2 - \beta r) F(r), \qquad (3)$$

where $\alpha$ and $\beta$ are positive parameters whose values are to be determined in terms of potential parameters a, b, c and d.

Substituting by Eq. (3) into Eq. (4), we obtain

$$\left[\frac{d^2}{dr^2} + \left(\frac{N-1}{r} - 4\alpha r - 2\beta\right)\frac{d}{dr} + (4\alpha^2 - 2\mu a)r^2 + (4\alpha\beta - 2\mu b)r + \left(2\mu c - \beta(N-1)\right)\frac{1}{r} - l(l+N-2))\frac{1}{r^2} + (\beta^2 + 2\mu E - 2\alpha N)\right]F(r) = 0. \qquad (4)$$

Assume a series solution to the above equation

$$F(r) = \sum_{n=0}^{\infty} a_n \, r^{\frac{3n}{2}+l}, \qquad (5)$$

where $a_n$ are expansion coefficients to be determined later. It is to be noted that the basic purpose of choosing a power series like $r^{\frac{3n}{2}+l}$ in the above expression is to avoid degeneracy in energy eigenvalues for some states considered . By substituting Eq. (5) into Eq. (4) and collection powers of r, we obtain

$$\sum_{n=0}^{\infty} a_n \left[ ((3n+2l)(3n+2l+2N-4) - 4l(l+N-2) - 8\mu d)r^{\frac{3n}{2}+l-2} + 4((2\mu c - \beta(N-1) - \beta(3n+2l))r^{\frac{3n}{2}+l-1} + 4((\beta^2 + 2\mu E - 2\alpha N - 2\alpha(3n+2l))r^{\frac{3n}{2}+l} + 8(2\alpha\beta - \mu b)r^{\frac{3n}{2}+l+1} + 8(2\alpha^2 - \mu a)r^{\frac{3n}{2}+l+2} \right] = 0. \quad (6)$$

This equation, after equating each coefficient of r to zero gives the following relations:

$$E = \frac{2\alpha(3n+2l+N) - \beta^2}{2\mu}, \quad (7)$$

$$a = \frac{2\alpha^2}{\mu}, \quad (8)$$

$$b = \frac{2\alpha\beta}{\mu}, \quad (9)$$

$$c = \frac{\beta(N+3n+2l-1)}{2\mu}, \quad (10)$$

$$(3n+2l)(3n+2l+2N-4) - 4l(l+N-2) - 8\mu d = 0. \quad (11)$$

The anstaz parameters $\alpha$ and $\beta$ may be obtained from equations (8) and (9), respectively as

$$\alpha = \sqrt{\frac{a\mu}{2}}, \qquad a > 0 \quad (12)$$

$$\beta = b\sqrt{\frac{\mu}{2a}}, \qquad a > 0 \quad (13)$$

thus, the energy eigenvalue becomes

$$E = \sqrt{\frac{a}{2\mu}}(N + 3n + 2l) - \frac{b^2}{4a} \quad (14)$$

## 3. Mass spectra of heavy and heavy-light mesons

In this section, we calculate spectra of the heavy quarkonium and heavy-light mesons that have quark and antiquark flavor, the mass of quarkonium is calculated in the 3-dimensional (N=3), so we apply the following relation

$$M = m_q + m_{\bar{q}} + E_{nl}, \tag{15}$$

where m is bare quark mass for quarkonium. By using Eq. (14), we can write Eq. (15) as follows:

$$M = m_q + m_{\bar{q}} + \sqrt{\frac{a}{m_c}}(3 + 3n + 2l) - \frac{b^2}{4a}. \tag{16}$$

In literature, the masses of charm and bottom quarks are taken between 1.2 GeV to 1.8 GeV and 4.8 GeV to 5.3 GeV, respectively. [24, 25, 2, 26 − 28]. In this work, we have chosen $m_c = 1.48\ GeV, m_b = 4.823\ GeV, m_s = 0.419\ GeV$, and $m_d = m_u = 0.220$ GeV [ 29 − 31]. The potential parameters a and b for various mesons are determined using Eq. (16). In case of charmonium, the values of a and b are calculated by solving two algebraic equations in a and b, which are obtained by inserting experimental values of M for 2S, 2P in Eq. (16). In case of bottomonium, the values of a and b are calculated by solving two algebraic equations in a and b, which are obtained by inserting experimental values of M for 1S, 2S in Eq. (16). In case of $\bar{b}c$, the values of a and b are calculated by solving two algebraic equations in a and b, which are obtained by using experimental values of M for 1S, 2S in Eq. (16). Similarly, for $c\bar{s}, b\bar{s}$ and $b\bar{q}$, parameters a and b are determined. The value of parameter d was calculated by using Eq. (11).

## 4. Results and Discussion

Kumar and Chand [6] calculated the mass spectra of $c\bar{c}$ and $b\bar{b}$ within series solution to the N- dimensional radial schrödinger equation for the quark– antiquark interaction potential. The quark-antiquark interaction potential, which consists of harmonic, linear and coulomb potential terms. There is a defect in this research.

1- The following is derived in Ref. [6]

$$(3n+2l)(3n+2l+2N-4)-4l(l+N-2) = 0. \tag{17}$$

This equation was not valid for all n = 0, 1 … this equation acts as constraints on the given system with l≥0 and n≥ 0. However, the only acceptable solution to this equation is n = 0. To overcome on this difficulty, we add the term $\frac{d}{r^2}$ to Cornell plus harmonic potential that displayed in Eq. (1). We obtained the following relation that has no constraints on l≥0 and n≥ 0

$$(3n + 2l)(3n + 2l + 2N - 4) - 4l(l + N - 2) - 8\mu d = 0. \qquad (18)$$

**2**-Kumar and Chand [6] calculated only the mass spectra of $c\bar{c}$ and $b\bar{b}$. Thus, we extend our calculation to heavy-light meson such as $c\bar{s}, \bar{b}c, b\bar{s}$ and $b\bar{q}$ and as we see in the following discussion.

In Table (1), we obtained the mass of charmonium by using Eq. (16), where $m_c = $ 1.48 GeV. In this table, we note that the present results are improved in comparison with recent Refs. [1, 5, 23, 32] and are agreement in comparison with the experimental data. In this work, we obtained total error for charmonium mass 0.162 as displayed in Table (1). In Ref. [5], the SE is solved by using the asymptotic iteration method and also they used Cornell potential. They obtained the total error for charmonium 0.441. In Ref. [23], the SE is solved by using the Nikiforov-Uvarov method with employing the extended Cornell potential which is a particular case from the present potential at d = 0. A similar situation in Ref. [24] that the authors obtained the total error for charmonium equals 0.905. In Ref. [32], the SE is solved by using the Nikiforov-Uvarov method and the authors obtained total error for charmonium 0.304.

In Table (2), we obtained the mass of bottomonium by using Eq. (16), where $m_b$=4.623 GeV. In this table, we note that the present results are improved in comparison with recent Refs. [1, 5, 6, 23]. We obtained the total error 0.074 for bottomonium. In recent Ref. [5], authors obtained the total error for bottomonium 0.172. In Ref [23], authors obtained the total error for bottomonium 0.485. In Ref. [6], the SE is solved by using series method with employing the extended Cornell potential. They obtained the total error for bottomonium equals 0.090. In Ref. [1], authors obtained total error for bottomonium equals 0.075.

In Table (3), we obtained the mass of $\bar{b}c$ by using Eq. (16). Where $m_b = 4.823$ GeV and $m_c = 1.209$ GeV, we note that the present results are improved in comparison with recent Refs. [1, 5, 25, 30, 34 ]. The present result for 1S state of $\bar{b}c$ closes with experimental data. In the Ref. [34], authors calculated the mass spectra of $\bar{b}c$ with non-relativistic treatment for $Q\bar{Q}$ systems and they have considered a general power potential color Coulomb term. They obtained to total error for $\bar{b}c$ about 0.01. In Ref. [30], authors calculated the mass spectra of $\bar{b}c$ using relativistic quark model based on quasipotential approach. They obtained to total error for $\bar{b}c$ 0.0008.

In Table (4), we obtained the mass of $c\bar{s}$ by using Eq. (16). Where $m_c = 1.209$ GeV and $m_s = 0.419$ GeV. We note that the present results are improved in comparison with recent Refs. [1, 3, 6, 31]. In the Ref. [31], authors obtained total error for $c\bar{s}$ 0.0108. In the present calculations, obtained results close with experimental data.

In Table (5), we obtained the mass of $b\bar{s}$ by using Eq. (16), where $m_b = 4.823$ GeV and $m_s = 0.419$ GeV, we note that the present results are improved in comparison with recent Ref. [29]. The authors obtained total error for $b\bar{s}$ 0.011. In the present calculation, we obtained total error 0.00001 and we note that 1S and 1P states close with experimental data. In Table (6), we obtained the mass of $b\bar{q}$ by using Eq. (16), where $m_b = 4.823$ GeV and $m_d = m_u = 0.220$ GeV. We note that the present results are improved in comparison with recent Ref. [29]. In Ref. [29], they obtained total error 0.019. In the present work, total error 0.00003 and we note that 1P state closes with experimental data.

In this work, one interests to study the effect on the dimensional number on quarkonium mass. The motivation for this as a natural consequence of the unification of the two modern theories of quantum mechanics and relativity and the emergence of the string theory, the investigation of the Standard Model particles in extra or higher-dimensional space is a hot topic of interest. From the experimental point of view, the investigation of the existence of extra dimensions is one of the primary goals of the LHC. The search for extra dimensions with the ATLAS and CMS detectors is discussed in Ref. [37]. In Tables (1-6), we note that quarkonium mass increases with increasing dimensionality number when we take N = 5. This means that that binding energy increases with increasing dimensionality number. The investigation of quarkonium in the higher dimensional space has taken attention in recent works and the dimensionality number plays an important role in changing the binding energy and dissociation temperatures as in Refs. [38, 39]. Also, in Ref. [40], the authors showed that the dimensionality number plays an important role for applying of the limitation of non-relativistic models and show also that the quarkonium mass increases with increasing the dimensionality number.

**Table 1:** Mass spectra of charmonium in (GeV) (a = 0.058 $GeV^3$, b = 0.3366 $GeV^2$)

| States | C | P. W. | [1] | [5] | [23] | [32] | N = 5 | Exp.[33] |
|---|---|---|---|---|---|---|---|---|
| 1S | 1.1486 | 3.068 | 3.096 | 3.096 | 3.078 | 3.096 | 3.461 | 3.068 |
| 1P | 2.2972 | 3.464 | 3.259 | 3.214 | 3.415 | 3.255 | 3.857 | 3.525 |
| 1D | 3.4459 | 3.861 | 3.511 | 3.412 | 3.752 | 3.504 | 4.253 | 3.770 |
| 2S | 2.8716 | 3.663 | 3.686 | 3.686 | 4.187 | 3.686 | 4.055 | 3.663 |
| 2P | 4.0202 | 4.059 | 3.779 | 3.773 | 4.143 | 3.779 | 4.451 | - |
| 3S | 4.5945 | 4.258 | 4.037 | 4.275 | 5.297 | 4.040 | 4.649 | 4.159 |
| 4S | 6.3175 | 4.852 | 4.263 | 4.865 | 6.407 | 4.269 | 5.243 | 4.421 |
| total | | 0.162 | 0.226 | 0.441 | 0.905 | 0.304 | | |

**Table 2:** Mass spectra of bottomonium in (GeV) (a = 0.1698 $GeV^3$, b= 0.7131 $GeV^2$)

| States | C | P. W. | [1] | [5] | [23] | [6] | N = 5 | Exp. [33] |
|---|---|---|---|---|---|---|---|---|
| 1S | 0.7878 | 9.460 | 9.460 | 9.460 | 9.510 | 9.510 | 9.835 | 9.460 |
| 1P | 1.5757 | 9.8354 | 9.612 | 9.492 | 9.862 | 9.862 | 10.210 | 9.900 |
| 1D | 2.3636 | 10.210 | 9.849 | 9.551 | 10.214 | 10.214 | 10.586 | 10.161 |
| 2S | 1.9697 | 10.023 | 10.023 | 10.023 | 10.627 | 10.038 | 10.398 | 10.023 |
| 2P | 2.7576 | 10.398 | 10.111 | 10.038 | 10.944 | 10.390 | 10.773 | 10.260 |
| 3S | 3.1515 | 10.585 | 10.361 | 10.585 | 11.726 | 10.566 | 10.961 | 10.355 |
| 4S | 4.3334 | 11.148 | 10.580 | 11.148 | 12.834 | 11.094 | 11.524 | 10.580 |
| Total error | | 0.074 | 0.075 | 0.172 | 0.485 | 0.090 | | |

**Table 3:** Mass spectra of $\bar{b}c$ in (GeV) (a = 0.2281 $GeV^3$, b = 2.410 $GeV^2$)

| States | c | P. W | [5] | [1] | [25] | [34] | [30] | N=5 | Exp. [35] |
|---|---|---|---|---|---|---|---|---|---|
| 1S | 2.594 | 6.277 | 6.277 | 6.277 | 6.270 | 6.349 | 6.332 | 6.670 | 6.277 |
| 1P | 7.258 | 6.4234 | 6.340 | 6.666 | 6.699 | 6.715 | 6.734 | 7.059 | - |
| 1D | 10.888 | 6.569 | 6.452 | - | - | - | 7.072 | 7.448 | - |
| 2S | 9.073 | 6.4963 | 6.814 | 7.042 | 6.835 | 6.821 | 6.881 | 7.253 | - |
| 2P | 12.702 | 6.6419 | 6.851 | 7.207 | 7.091 | 7.102 | 7.126 | 7.642 | - |
| 3S | 14.517 | 6.7148 | 7.351 | 7.384 | 7.193 | 7.175 | 7.235 | 7.837 | - |
| 4S | 19.961 | 6.9333 | 7.889 | - | - | - | - | 8.420 | - |
| Total | | - | - | - | 0.0001 | 0.01 | 0.0008 | | |

**Table 4:** Mass spectra of $c\bar{s}$ in (GeV) (a = 0.036 $GeV^3$, b = 0.4549 $GeV^2$)

| States | c | P. W | [31] | [3] | [6] | [1] | N = 5 | Exp. [27] |
|---|---|---|---|---|---|---|---|---|
| 1S | 3.038 | 2.258 | 2.129 | 1.968 | 2.512 | 1.968 | 2.562 | - |
| 1P | 6.077 | 2.558 | 2.549 | 2.565 | 2.649 | 2.566 | 2.859 | - |
| 1D | 9.115 | 2.859 | 2.899 | 2.857 | 2.859 | 2.857 | 3.156 | 2.859 |
| 2S | 7.596 | 2.709 | 2.732 | 2.709 | 2.709 | 2.709 | 3.008 | 2.709 |
| 2P | 10.635 | 3.009 | 3.018 | - | 2.860 | - | 3.305 | - |
| 3S | 12.154 | 3.159 | 3.193 | 2.932 | 2.906 | 2.932 | 3.454 | - |
| 4S | 16.712 | 3.609 | 3.575 | - | 3.102 | - | 3.900 | - |
| Total | | - | .0108 | 0.00006 | - | 0.00003 | | |

**Table 5:** Mass spectra of $b\bar{s}$ in (GeV) (a = 0.255 GeV$^3$, b = 2.26 GeV$^2$)

| States | C | Present work | [29] | N =5 | Exp.[35 − 36] |
|---|---|---|---|---|---|
| 1S | 5.098 | 5.416 | 5.450 | 6.579 | 5.415 |
| 1P | 10.197 | 5.830 | 5.857 | 7.020 | 5.830 |
| 1D | 15.296 | 6.245 | 6.182 | 7.461 | - |
| 2S | 12.746 | 6.038 | 6.012 | 7.241 | - |
| 2P | 17.845 | 6.452 | 6.279 | 7.682 | - |
| 3S | 20.394 | 6.659 | 6.429 | 7.902 | - |
| 4S | 28.042 | 7.281 | 6.773 | 8.564 | - |
| Total | | 0.00001 | 0.011 | | |

**Table 6:** Mass spectra of $b\bar{q}$ in $(GeV)$ (a = 0.1997 GeV$^3$, b = 2.068 GeV$^2$).

| States | C | P. W. | [29] | N = 5 | Exp.[35 − 36] |
|---|---|---|---|---|---|
| 1S | 7.095 | 5.326 | 5.371 | 5.727 | 5.325 |
| 1P | 14.190 | 5.724 | 5.777 | 6.125 | 5.723 |
| 1D | 21.285 | 6.122 | 6.110 | 6.523 | - |
| 2S | 17.738 | 5.923 | 5.933 | 6.324 | - |
| 2P | 24.833 | 6.321 | 6.197 | 6.722 | - |
| 3S | 28.380 | 6.520 | 6.355 | 6.921 | - |
| 4S | 39.023 | 7.117 | 6.703 | 7.518 | - |
| Total | | 0.00003 | 0.019 | | |

# 5-Summary and conclusion

In the present work, we calculated the energy eigenvalues in the N-dimensional form by solving the N- radial Schr$\ddot{o}$dinger equation using power series technique. This method plays an important role in solving SE. In Ref. [6], the series method is used to solve the SE. In this work, we treated the weakness points that found in Ref. [6] by suggested a new form of quark-antiquark interaction that is displayed by Eq. (1). Also, we extended the calculations of Ref. [6] to include the heavy-light meson. The present results are improved in comparison with other recent works such as [5, 6, 23, 25, 29 – 32, 34] and in good agreement with experimental data. In addition, we study the effect of dimensionality number on quarkonium masses which is not considered in the previous works. We hope to extent this work to include the effect of medium on quarkonium properties as a future work.

# 6-References